\documentclass[]{interact}

\usepackage{epstopdf}
\usepackage[caption=false]{subfig}

\usepackage[numbers,sort&compress]{natbib}
\usepackage{amsmath,amssymb,amsfonts}
\usepackage{mathrsfs}
\usepackage{chemformula}
\usepackage{color}
\usepackage[normalem]{ulem}
\usepackage{stmaryrd}

\usepackage[parfill]{parskip} 
\graphicspath{{./figures/}}

\bibpunct[, ]{[}{]}{,}{n}{,}{,}
\renewcommand\bibfont{\fontsize{10}{12}\selectfont}

\theoremstyle{plain}

\theoremstyle{definition}

\theoremstyle{remark}

\newcommand{\red}[1]{\textcolor{black}{#1}}
\newcommand{\blue}[1]{\textcolor{black}{#1}}

\newcommand{\beq}{\begin{equation}}
\newcommand{\eeq}{\end{equation}}
\newcommand{\pfr}[2]{\ensuremath{\frac{\partial #1}{\partial #2}}}
\newcommand{\pfi}[2]{\ensuremath{{\partial #1}/{\partial #2}}}
\newcommand{\ep}{\epsilon}

\newcommand\Pec{\mbox{\textit{Pe}}}
\newcommand\Pra{\mbox{\textit{Pr}}}
\newcommand\Lew{\mbox{\textit{Le}}}

\newcommand\Rey{\mbox{\textit{Re}}}
\newcommand{\vect}[1]{\mathbf{#1}}

\begin{document}

\articletype{Research article}

\title{A thick reaction zone model \blue{for premixed flames in two-dimensional channels }}

\author{
\name{Prabakaran Rajamanickam and Joel Daou}
\affil{Department of Mathematics, University of Manchester, Manchester, UK}
}

\maketitle

\begin{abstract}
 Direct interactions between the flow field and the chemical reaction in premixed flames occur when the reaction zone thickness is comparable to, or greater than flow length scales. To study such interactions, a laminar model is considered that has direct bearings to steadily propagating deflagrations in a Hele-Shaw channel with a background plane Poiseuille flow. The study employs asymptotic analyses, pertaining to large activation energy and lubrication theories and considers a distinguished limit where the channel width is comparable to the reaction zone thickness, with account being taken of thermal-expansion \blue{and heat-loss} effects. The reaction zone structure and burning rates depend on three parameters, namely, the Peclet number, $\mathcal{P}$, the Lewis number, $\Lew$ and the ratio of channel half-width to reaction zone thickness, $\lambda_*$. In particular, when the parameter $\lambda_*$ is small wherein the reaction zone is thick, transport processes are found to be controlled by Taylor's dispersion mechanism and an explicit formula for the effective burning speed $S_T$ is obtained. The formula indicates that $S_T/S_L \propto 1/\Lew$ for $\mathcal{P}\gg 1$, which interestingly coincides with a recent experimental prediction of the turbulent flame speed in a highly turbulent jet flame. The results suggest that the role played by differential diffusion effects is significant both in the laminar and turbulent cases. The reason for the peculiar $1/\Lew$ dependence can be attributed, at least in our laminar model, to Taylor dispersion. Presumably, this dependence may be attributed to a similar but more general mechanism in the turbulent distributed reaction zone regime, rather than to diffusive-thermal curvature effects. The latter effects play however an important role in determining the effective propagation speed for thinner reaction zones, in particular, when $\lambda_*$ is large in our model. \blue{It is found that the magnitude of heat losses at extinction, which directly affects the mixture flammability limits, is multiplied by a factor $1/\Lew^2$ in comparison with those corresponding to the no-flow case in narrow channels.}
\end{abstract}


\section{Introduction}

In a number of practical problems in premixed combustion, laminar flames may be regarded as hydrodynamic discontinuities provided their thickness $\delta_L$ is small compared with a characteristic length scale $L$ of the flow. They may also be regarded as reaction sheet discontinuities flanked by preheat and post-flame zones, when attention is needed for scales greater than the sheet thickness $\delta_L/\beta$, but of the order of $\delta_L$, where $\beta$ represents the Zeldovich number. Such treatments of laminar flames proved to be fruitful in developing an understanding of their behaviours; for example hydrodynamic instabilities and diffusive-thermal instabilities are \red{typically investigated using these two approximations}, respectively. There are situations, however, where neither the flame nor its reaction zone may be considered thin. A notable area where this may be the case is that of turbulent combustion in which a wide range of flow scales exists. More specifically, the thin flame or thin reaction sheet approximations may not be helpful to elucidate how the flame structure is influenced by the small scales in a turbulent flow. To gain insight into thick reaction zone regimes in such or similar situations, a laminar model problem is considered in this paper.

Flames constricted to narrow channels \red{have been} demonstrated to be convenient platforms for this type of studies both from the theoretical~\cite{short2009asymptotic,pearce2014taylor,daou2018taylor,daou2001flame,daou2002influence,kurdyumov2011lewis,fernandez2018analysis} and experimental~\cite{almarcha2015experimental,wongwiwat2015flame,sarraf2018quantitative,veiga2019experimental} point of view, as they can shorten the transverse length scale considerably. At small transverse length scales, combustion is restricted due to excessive heat losses at the channel walls, although such severe limitations do not occur in turbulent combustion. Thus setting aside the practical difficulties, it is still worthwhile to study strongly confined flames as a means for exploring strongly sheared reaction layers. The extent of flame confinement and shear effects are controlled by two independent parameters, namely\footnote{We let the modified Peclet number $\mathcal{P}$ and the other parameter $\lambda_*$ introduced in the abstract to emerge naturally through the analysis provided in $\S 3$ and $\S 5$.}
\begin{equation}
    \ep = \frac{L}{\delta_L}, \qquad  \qquad \Pec = \frac{U_mL}{D_{T,u}}    \nonumber
\end{equation}
where $L$ denotes the channel half-width, $U_m$ the mean velocity of the flow, $D_{T,u}$ the thermal diffusivity in the unburnt mixture, $\ep$ the channel half-width scaled by the flame thickness and $\Pec$ the Peclet number, numerically of the order of the Reynolds number for gaseous flows. In terms of these two parameters, it can be seen that strong confinement is achieved for $\ep \ll 1$, whereas effects of shear convection will be significant if $\Pec\sim O(1)$ or larger.

\begin{figure}[h!]
\centering
\includegraphics[width=1\textwidth]{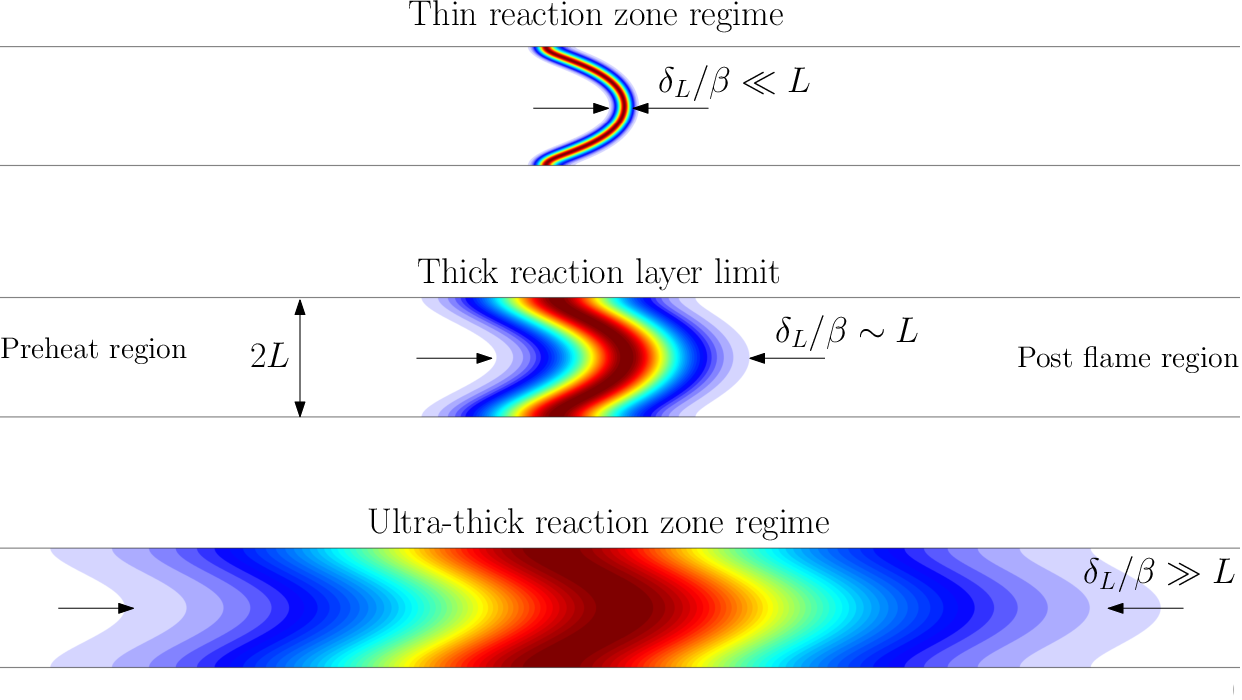}
\caption{Typical reaction rate contours of flames propagating in a channel flow.}
\label{fig:schematic}
\end{figure}

In general, by comparing the thickness of the reaction zone $\delta_L/\beta$ with the channel half-width $L$, we can classify the combustion modes into three different regimes (see Fig.~\ref{fig:schematic}):
\begin{itemize}
    \item \textit{Thin reaction zone regime} ($\ep\gg 1/\beta$): In this regime, the chemical reaction is confined to a thin reaction sheet with thickness $\delta_L/\beta$. This regime can be further classified into, thin flames where $\ep\gg 1$ and moderately thick flames where $\ep\sim 1$. 
    \item \textit{Thick reaction layer limit} ($\ep\sim 1/\beta$): In this limit, the  reaction layer is as thick as the channel width, or the flow scale. 
    \item \textit{Ultra-thick reaction zone regime} ($\ep\ll 1/\beta$): Here the reaction layer width is very large in comparison with the flow scale.
\end{itemize}
Specific to the parallel flow geometry considered here with $\Pec\sim O(1)$, mixing of scalars are affected by Taylor's dispersion mechanism, as noted in~\cite{pearce2014taylor,daou2018taylor} for premixed flames and~\cite{linan2020taylor,rajamanickam2022effects} for diffusion flames. Taylor's dispersion mechanism on account of the interaction between molecular diffusion and shear convection occurring on a length scale $x\sim L$ manifests as an apparent diffusion process for scales $x\gg L$. It is evident from Fig.~\ref{fig:schematic} that the dispersion mechanism is applicable to preheat and post-flame zones but not the reaction zone in the thick reaction layer limit and to all zones in the ultra-thick reaction zone regimes; no Taylor's dispersion mechanism manifests for the thin reaction zone regimes. As the objective of this paper is to investigate the specific interactions mentioned above, it thus becomes necessary to consider the distinguished limit $\epsilon \sim 1/\beta \ll 1$ with $\Pec\sim O(1)$.

It should be noted that apart from studies adapting the condition $\Pec\sim O(1)$ such as \cite{short2009asymptotic,pearce2014taylor,daou2018taylor}, many analytical investigations~\cite{daou2001flame,daou2002influence,kurdyumov2011lewis} gave attention to the limit $A=U_m/S_L\sim O(1)$ where $A$ denotes the flow amplitude and $S_L$ the laminar burning speed of a planar flame with which we can define $\delta_L=D_{T,u}/S_L$. The relation between $A$ and $\Pec$ is given by $\Pec = \epsilon A$ indicating that the correspondence between $\Pec$ and $A$ should be made according to the flame type described above. In particular, $\Pec\sim O(1)$ does not always imply significant interactions between shear and diffusive transport of the flame. For instance, in thin flames, if $\Pec\sim O(1)$ then $A\ll 1$ which implies that the flow can wrinkle the flame surface, but not affect its structure. This once again emphasizes the need for the aforementioned double limit.

With the objective of exploring thick reaction layer flamelets modelled with a strongly confined, strongly sheared flames, the study considers the distinguished limit $\ep \sim 1/\beta \ll 1$ with  $\Pec\sim O(1)$. The study is an extension of~\cite{pearce2014taylor,daou2018taylor} with specific focus given to thick reaction layers. Formulation of the problem is given in the next section, whereas $\S 3$ and $\S 4$ provide the analysis and numerical results for the limits mentioned above. In $\S$5, implications of the model problem in highly turbulent flames are discussed. Finally, conclusions are provided in $\S 6$.

\section{Formulation of the problem (steadily propagating, nonadiabatic premixed flame in a Hele-Shaw burner)}
The longitudinal velocity profile of a fully developed flow containing a reacting gas mixture in a Hele-Shaw channel driven by a constant pressure gradient is given by 
\begin{equation}
    \frac{3U_m}{2}\left(1-\frac{y^2}{L^2}\right)    \nonumber
\end{equation}
where $y$ denotes the transverse coordinate. The flow direction is from left to right if $U_m>0$ and right to left otherwise. Consider a propagating deflagration in this flow field, moving at a constant speed $U$ from right to left. The propagation speed $U$ comprises two parts, namely, $U=-U_m + S_T$, where the first one arises because the mixing layer convects to the right (left) with the mean velocity $U_m>0$ ($U_m<0$) and the other, $S_T$, is the effective burning speed \blue{which represents the flame speed relative to the mean flow}. To render the problem steady, we shall consider a Cartesian frame $(x,y,z)$ with velocity components $(v_x,v_y,v_z)$ attached to the steadily propagating deflagration. In this frame, as $x\rightarrow -\infty$, the velocity profile takes the form
\begin{equation}
    v_x=\frac{U_m}{2}\left(1-\frac{3y^2}{L^2}\right) + S_T, \qquad v_y=0, \qquad v_z = 0 \label{UT}
\end{equation}
where the first term $U_m(1-3y^2/L^2)/2$ in the expression for $v_x$ represents the velocity fluctuation from its mean value. This term will be seen below to play a vital role in determining the reaction layer structure and $S_T$.

For simplicity, we shall adapt a single-step irreversible Arrhenius reaction with large activation energies to describe the chemistry in a reacting mixture assumed to be fuel lean. For sufficiently fuel-lean conditions, the mass of fuel burned per unit volume per unit time is then given by $\rho B Y_F e^{-E/RT}$, which involves the pre-exponential factor $B$, the gas density $\rho$, the fuel mass fraction $Y_F$, the temperature $T$, the activation energy $E$ and the universal gas constant $R$. The corresponding adiabatic flame temperature $T_{ad}$, heat release parameter $\alpha$ and Zeldovich number $\beta$ are defined as
\begin{equation}
    T_{ad} = T_u + \frac{qY_{F,u}}{c_p}, \qquad \alpha = \frac{T_{ad}-T_u}{T_{ad}}, \qquad \beta = \frac{E}{RT_{ad}}\frac{T_{ad}-T_u}{T_{ad}}     \nonumber
\end{equation}
where $q$ denotes the amount of energy released by the chemical reaction per unit mass of fuel consumed and $c_p$ is the specific heat at constant pressure, assumed to be a constant. Throughout this document, the subscript $u$ refers to unburnt gas mixture. In the limit $\beta\rightarrow \infty$, the laminar burning speed of a one-dimensional planar flame has the following expression
\begin{equation}
    S_L =  \left[2\Lew \beta^{-2} B  D_{T,u}(1-\alpha)^2 e^{-E/RT_{ad}}\right]^{1/2}   \label{SL} 
\end{equation}
where $\Lew$ is the Lewis number of the fuel. The above formula will be used to normalize the gas velocity components and the burning speed, $S_T$, of the flame.

Conductive heat losses at the wall can be included by imposing the boundary condition $\pfi{T}{y}=-K(T-T_u)$ at $y=\pm L$ with $K$ denoting the heat transfer coefficient of the wall. For narrow channels, the appropriate scaling for flame quenching to heat loss is $\beta LK\sim \ep^2$~\cite{daou2002influence} and thus we introduce a nondimensional heat loss parameter
\begin{equation}
   \kappa = \frac{\beta LK}{\ep^2}.    \nonumber      
\end{equation}

The analysis is facilitated by introducing the following non-dimensional variables
\begin{align}
   & (\xi,\eta) = \frac{1}{\delta_L}(x,y/\ep), \quad (u+V,v) =\frac{1}{S_L} (v_x,v_y), \quad p = \frac{P}{\rho_u S_L^2\Pra},\nonumber \\ &y_F = \frac{Y_F}{Y_{F,u}}, \quad \theta = \frac{T-T_u}{T_{ad}-T_u}, \quad \varrho=\frac{\rho}{\rho_u},  \quad V=\frac{S_T}{S_L}      \nonumber
\end{align}
where $P$ denotes the \blue{modified} pressure, \blue{which accounts for the hydrostatic pressure and includes a term proportional to the divergence of the velocity field}, $\Pra$ the Prandtl number \blue{and $V$  the effective burning speed scaled by $S_L$.} For low Mach number flows, the equation of state simplifies to $\rho T=\rho_u T_u$, whereas for the diffusion coefficients, we shall assume a linear dependence on the temperature so that $\rho \nu = \rho_u\nu_u$, $\rho D_F=\rho_u D_{F,u}$ and $\rho D_T=\rho_u D_{T,u}$. The required equations then reduce to
\begin{align}
   \pfr{[\varrho (u+V)]}{\xi} + \frac{1}{\ep}\pfr{(\varrho v)}{\eta}&=0,  \label{gov1} \\
    \frac{\varrho}{\Pra} \left[(u+V) \pfr{}{\xi} + \frac{v}{\ep}\pfr{}{\eta}\right]u &= - \pfr{p}{\xi} +  \pfr{^2u}{\xi^2} + \frac{1}{\ep^2}\pfr{^2u}{\eta^2}, \label{gov2}\\
    \frac{\varrho}{\Pra} \left[(u+V) \pfr{}{\xi} + \frac{v}{\ep}\pfr{}{\eta}\right]v &= - \frac{1}{\ep}\pfr{p}{\eta} + \pfr{^2v}{\xi^2} + \frac{1}{\ep^2}\pfr{^2v}{\eta^2},   \label{gov3}\\
    \varrho \left[(u+V) \pfr{}{\xi} + \frac{v}{\ep}\pfr{}{\eta}\right]y_F &= \frac{1}{\Lew} \left(\pfr{^2y_F}{\xi^2} + \frac{1}{\ep^2}\pfr{^2y_F}{\eta^2}\right) -  \omega, \label{gov4}\\
   \varrho \left[(u+V) \pfr{}{\xi} + \frac{v}{\ep}\pfr{}{\eta}\right]\theta &=  \pfr{^2\theta}{\xi^2} + \frac{1}{\ep^2}\pfr{^2\theta}{\eta^2} +  \omega ,    \label{gov5}\\
    \varrho \left(1+ \frac{\alpha \theta}{1-\alpha}\right) &= 1  \label{gov6}
\end{align}
where 
\begin{equation}
    \omega = \frac{\beta^2\varrho y_F}{2\Lew (1-\alpha)}\exp\left[\frac{-\beta(1-\theta)}{1-\alpha(1-\theta)}\right].  \nonumber
\end{equation}
Assuming symmetry in the $y$-direction, the problem can be solved in the upper-half infinite strip, that is to say, for $\eta\in[0,1]$ and $\xi\in(-\infty,\infty)$. The required boundary conditions for this domain are given by
\begin{align}
    u - \frac{\Pec}{2\ep}(1-3\eta^2)=v= y_F-1= \theta=0,\qquad \text{as} \qquad \xi\rightarrow -\infty, \label{BC1}\\
    \pfr{u}{\xi}=\pfr{v}{\xi}=y_F=\pfr{\theta}{\xi}=0, \qquad \text{as} \qquad \xi\rightarrow \infty, \label{BC2}\\
    \pfr{u}{\eta}=v=\pfr{y_F}{\eta}=\pfr{\theta}{\eta}=0, \qquad \text{at} \qquad \eta=0, \label{BC3}\\
    u+\frac{\Pec}{\ep}=v=\pfr{y_F}{\eta}=\pfr{\theta}{\eta}+\ep^2\frac{\kappa\theta}{\beta}=0, \qquad \text{at} \qquad \eta=1. \label{BC4}
\end{align}

The problem proposed here can be integrated numerically, in fact, for arbitrary values of $\ep$ as has been done, for example, in~\cite{daou2018taylor} in the adiabatic ($\kappa=0$) case. As mentioned in $\S$1, this paper is devoted to the consideration of thick reaction zones where $\ep\sim 1/\beta$, accompanied by the limit $\ep\rightarrow 0$. The study of this distinguished limit is achieved by introducing an order-unity parameter 
\begin{equation}
    \lambda = \beta\ep.    \label{lambda}
\end{equation}
Investigation of $\lambda\sim 1$ cases i.e., those corresponding to thick reaction layers, will also illuminate the asymptotic behaviors of cases where $\lambda\ll 1$ corresponding to the ultra-thick reaction zone regimes and $\lambda\gg 1$ corresponding to the moderately thick flames with a thin reaction zone. Since $\beta=\lambda/\epsilon$, statements made in terms of $\beta$ to identify different flame layers, will be made now in terms of $\lambda/\ep$.

\section{Analysis for Taylor-dispersion controlled flames}
In the distinguished limit $\epsilon \rightarrow 0$ with $\lambda\equiv \beta\epsilon \sim O(1)$, the reaction layer longitudinal extent is of the order of channel half-width, $L$, which in non-dimensional terms read $\xi \sim \ep$, if we assume, without loss of generality, that the reaction zone lies in the neighbourhood of the point $\xi=0$. The entire flame structure has three regions: a convective-diffusive preheat zone with thickness of order $\xi \sim 1$, an inner reactive zone with thickness of order $\xi\sim \ep/\lambda$ and a post-flame zone with thickness of order $\xi \sim \lambda/\ep$ due to a convection heat-loss balance. As we will be interested only in the amount of downstream flux leaving the reaction layer, the post-flame zone can be studied for the region $\xi\sim 1$ where the temperature profile is linear. Thus, both the preheat and the post-flame zone are analysed upstream and downstream of the plane $\xi=0$ with a longitudinal extent $\xi\sim 1$.

\subsection{Preheat and post flame zones}
The chemical reaction term $\omega$ appearing in~\eqref{gov4}-\eqref{gov5} is negligible at all orders both in the preheat and post flame zones because in the former the reacting mixture is chemically frozen and in the latter, chemical equilibrium has been reached.  Using $\ep$ as a small parameter, we introduce the following expansions
\begin{align}
    u = u_0/\ep + u_1 + \ep u_2 +\cdots, &\qquad  \qquad v = v_1 + \ep v_2 + \ep^2 v_3 + \cdots, \nonumber\\
    p = p_0/\ep^3 + p_1/\ep^2 + p_2/\ep + \cdots, &\qquad \qquad y_F = F_0 + \ep F_1 + \ep^2 F_2 + \cdots, \nonumber \\
    \theta= \theta_0 + \ep\theta_1 + \ep^2\theta_2 + \cdots, &\qquad \qquad   V= U_0 + \ep U_1 + \ep^2 U_2 + \cdots. \nonumber
\end{align}
The terms in the expansion for the density $\varrho = \varrho_0 + \ep\varrho_1 + \ep^2\varrho_2 + \cdots$ are given using~\eqref{gov6} by $\varrho_0 = 1/[1+\alpha\theta_0/(1-\alpha)]$, $\varrho_1=-\alpha\varrho_0^2\theta_1/(1-\alpha)$, etc. Substituting these perturbation series into the governing equations~\eqref{gov1}-\eqref{BC4}, one should be able to solve problems arising at each order. The analysis is quite similar to those presented in~\cite{pearce2014taylor,daou2018taylor,rajamanickam2022effects}, whence we do not pause here to provide a detailed account of the derivation, but just note that we arrive at the end to the following equations:
\begin{align}
    \pfr{p}{\xi} &= -\frac{3\Pec}{\ep^3} + \frac{1}{\ep^2}\left\{3 U_0 - \frac{9\gamma\Pec^2}{2 \Pra} \left[1 + \frac{2\Pra(1-\alpha)}{3\alpha}\right]\frac{d\varrho_0}{d\xi}\right\}+\cdots, \label{pexp}\\
    u &= \frac{\Pec}{2\ep}(1-3\eta^2) +  \left[\frac{1}{2}\frac{dp_1}{d\xi}(\eta^2-1) + \frac{\Pec^2}{\Pra}\frac{d\varrho_0}{d\xi}\left(\frac{\eta^4}{8}-\frac{\eta^6}{20}-\frac{3}{40}\right)\right] + \cdots, \label{uexp}\\
    v &= - \frac{\Pec}{2\varrho_0}\pfr{\varrho_0}{\xi}(\eta-\eta^3) + \cdots, \label{vexp}\\
    y_F &= F_0(\xi) + \ep \left[F_{1a}(\xi) + \varrho_0\Pec \Lew \frac{dF_0}{d\xi} \left(\frac{\eta^2}{4}-\frac{\eta^4}{8}\right)\right] + \cdots, \label{yfexp} \\
    \theta &= \theta_0(\xi) + \ep \left[\theta_{1a}(\xi) +\varrho_0\Pec \frac{d\theta_0}{d\xi} \left(\frac{\eta^2}{4}-\frac{\eta^4}{8}\right)\right] + \cdots . \label{texp}
\end{align}
Here, the functions $F_0(\xi)$ and $\theta_0(\xi)$ must satisfy the ordinary differential equations
\begin{align}
    U_0 \frac{dF_0}{d\xi} = \frac{1}{\Lew}\frac{d}{d\xi} \left[(1+\gamma \varrho_0^2\Pec^2\Lew^2)\frac{dF_0}{d\xi}\right] , \label{Fph}\\
    U_0 \frac{d\theta_0}{d\xi} = \frac{d}{d\xi} \left[(1+\gamma \varrho_0^2\Pec^2)\frac{d\theta_0}{d\xi}\right] - \frac{\ep\kappa\theta_0}{\lambda}, \label{thetaph}
\end{align}
where $\gamma=2/105$ and the parameters $1/\varrho_0+\gamma \varrho_0\Pec^2\Lew^2$ and $1/\varrho_0+\gamma \varrho_0\Pec^2$ represent the non-dimensional Taylor diffusion coefficients for fuel and temperature. If required, equations governing $F_{1a}(\xi)$ and $\theta_{1a}(\xi)$ can be derived from higher-order asymptotics.

Equations~\eqref{Fph}-\eqref{thetaph} are essentially the same as those that describe one-dimensional, nonadiabatic, planar flames, except for the modified diffusion coefficients. Much of what we know for these latter flames can be readily extended to our problem. The heat loss term in~\eqref{thetaph} lowers the flame temperature $\theta=\theta_f$ from its adiabatic value $\theta=1$ and the method of determining $\theta_f$ when $\epsilon/\lambda\ll 1$ is a standard one, see for instance~\cite{williams2018combustion} (pp.~271-276). With errors of order $\ep^2/\lambda^2$, we find
\begin{equation}
    \blue{\theta_f = 1 -  \frac{2\ep\kappa}{\lambda U_0^2}(1+\gamma\mathcal{P}^2) .}
\end{equation}
Here $\mathcal{P}=(1-\alpha)\Pec$ is simply the Peclet number based on thermal diffusivity at the adiabatic flame temperature\footnote{It will turn out that key parameter characterising the flame is not the cold temperature Peclet number $\Pec$, but the Peclet number $\mathcal{P}$ at the adiabatic flame temperature. It is indeed appropriate to formulate the original problem from the outset for the limits $\ep\rightarrow 0$ and $\mathcal{P}\sim 1$ since $\alpha\sim 1$.}. An effective Lewis number that accounts for flow induced effects can be defined as follows~\cite{daou2018taylor,linan2020taylor}
\begin{equation}
    \Lew_{\rm{eff}} = \Lew \frac{1+\gamma\mathcal{P}^2}{1+\gamma\mathcal{P}^2\Lew^2}. \label{Leeff}
\end{equation}

The fluxes entering/leaving the reaction sheet are obtained from a local analysis of the solutions of~\eqref{Fph} and~\eqref{thetaph}, satisfying the far-field boundary conditions. As $\xi\rightarrow 0^-$, we thus find
\begin{equation}
    \blue{\left.\frac{d\theta_0}{d\xi}\right |_{\xi=0^-} =  \frac{U_0}{1+\gamma\mathcal{P}^2} -  \frac{\ep\kappa}{\lambda U_0} , \qquad  \left.\frac{dF_0}{d\xi}\right |_{\xi=0^-} = -\frac{\Lew_{\rm{eff}}  U_0}{1+\gamma\mathcal{P}^2}} \label{outerflux1}
\end{equation}
whereas for the post flame zone, a similar observation as $\xi\rightarrow 0^+$ leads  to
\begin{equation}
    \left.\frac{d\theta_0}{d\xi}\right |_{\xi=0^+} = - \frac{\ep\kappa}{\lambda U_0}, \qquad  \left.\frac{dF_0}{d\xi}\right |_{\xi=0^+} = 0.  \label{outerflux2}
\end{equation}
These limiting behaviours will serve as matching conditions for the inner problem wherein $ U_0$ will be determined. Once $ U_0$ is solved, \eqref{Fph} and~\eqref{thetaph} can be solved numerically with appropriate boundary conditions to describe the preheat and post flame structure. If $\kappa=0$, these structures are identified implicitly from 
\begin{equation}
    F_0 = 1 - \exp\left(\int_0^\xi \frac{\Lew U_0 d\xi}{1+\gamma \varrho_0^2\Pec^2\Lew^2}\right), \qquad \theta_0 = \exp\left(\int_0^\xi \frac{ U_0 d\xi}{1+\gamma \varrho_0^2\Pec^2}\right)  \nonumber
\end{equation}
for the preheat zone and $F_0(\xi)=\theta_0(\xi)=0$ for the post-flame zone.

\subsection{Inner reaction layer problem}
To leading order, in the inner zone where $\xi\sim \epsilon/\lambda$, the density is constant and equal to $\varrho_{0}=1-\alpha$ and consequently the mass flux implied by~\eqref{uexp} and~\eqref{vexp} is given by
\begin{equation}
    \varrho u= \frac{\mathcal{P}}{2\ep}(1-3\eta^2) , \qquad \varrho v=0   \nonumber
\end{equation}
\blue{indicating that the flow is locally incompressible at leading order. Note that the parameter $\lambda=\beta \ep$ in~\eqref{lambda} represents the ratio of channel half-width to the reaction zone thickness of an one-dimensional, laminar, planar flame, that is to say, it uses the reaction-zone thickness definition $D_{T,u}/S_L\beta$. In Taylor dispersion controlled flames, a better definition of reaction zone thickness should include the flow-induced diffusion enhancement; the fluxes~\eqref{outerflux1} entering from the preheat zone are already seen to reflect the diffusion enhancement through $\mathcal{P}$. The appropriate definition of reaction zone thickness is $D_{T,u}(1+\gamma \mathcal{P}^2)/S_T\beta \sim D_{T,u}(1+\gamma \mathcal{P}^2)S_L/U_0\beta$. The ratio of channel half-width to the reaction zone thickness of Taylor dispersion-controlled flames then becomes}
\begin{equation}
    \blue{\lambda_* \equiv \frac{\beta \epsilon U_0}{1+\gamma \mathcal{P}^2}. \label{lambdaT}}
\end{equation}
\blue{In terms of $\xi$ coordinate the reaction layer thickness is $\xi \sim \ep/\lambda_*$, whereas the fuel mass fraction is $Y_F\sim 1/\beta=\epsilon/\lambda$ and the temperature departure from $\theta_f$ is $\theta_f-\theta \sim 1/\beta = \epsilon/\lambda$. Therefore, we introduce}
\begin{equation}
    \blue{X = \frac{\lambda_* \xi}{\ep} , \quad Y=  \frac{\lambda y_F}{\ep}, \quad \Theta = \frac{\lambda(\theta_f-\theta)}{\ep},  \quad  \Lambda =  U_0 \exp\left[\frac{ \kappa(1+\gamma\mathcal{P}^2)}{ U_0^2}\right] } \label{innerscale}
\end{equation}
\blue{in which the last parameter $\Lambda$, to be determined as an eigenvalue, provides the burning speed $U_0$ for a given value of $\kappa$. The parameter $\Lambda$ may be simply referred to as the \textit{adiabatic burning speed} since $U_0=\Lambda$ if $\kappa=0$.} Introducing these inner variables into~\eqref{gov1}-\eqref{BC4}, we obtain at leading order
\begin{align}
    \frac{\mathcal{P}}{2\lambda_*}(1-3\eta^2)\pfr{Y}{X} &= \frac{1}{\Lew}\left(\pfr{^2Y}{X^2} + \frac{1}{\lambda_*^2}\pfr{^2Y}{\eta^2}\right) - \blue{\frac{(1+\gamma\mathcal{P}^2)^2}{2\Lew\Lambda^2}Y e^{-\Theta} }  ,  \label{innerY} \\
   \frac{\mathcal{P}}{2\lambda_*}(1-3\eta^2)\pfr{\Theta}{X} &=\pfr{^2\Theta}{X^2} + \frac{1}{\lambda_*^2}\pfr{^2\Theta}{\eta^2} - \blue{\frac{(1+\gamma\mathcal{P}^2)^2}{2\Lew\Lambda^2}Y e^{-\Theta} } . \label{innerT}
\end{align}
 Matching with the conditions~\eqref{outerflux1}-\eqref{outerflux2} from the outer regions provide
\begin{align}
    \frac{1}{\Lew_{\rm{eff}}}\pfr{Y}{X}=\pfr{\Theta}{X}=- 1, \qquad &\text{as}\qquad X\rightarrow -\infty,   \label{innerbc1}\\
    \pfr{Y}{X} =\pfr{\Theta}{X}= 0, \qquad &\text{as} \qquad X\rightarrow \infty. \label{innerbc2}
\end{align}
At $\eta=0$ and $\eta=1$, we have
\begin{equation}
    \pfr{Y}{\eta}=\pfr{\Theta}{\eta}=0.  \label{innerbc3}
\end{equation}
The translational invariance in the $X$ direction can removed by imposing a finite positive value for $\Theta$ or $Y$ at a specified location, say at the origin. \blue{The problem just described is solved for given values of $\mathcal{P}$, $\Lew$ and $\lambda_*$ to obtain $\Lambda$, $Y(X,\eta)$ and $\Theta(X,\eta)$.}

Integrating the equations~\eqref{innerY} and~\eqref{innerT} across the channel utilizing~\eqref{innerbc3}, we obtain
\begin{align}
    \frac{dG_F}{dX} =  \frac{dG_T}{dX}=- \frac{(1+\gamma\mathcal{P}^2)^2}{2\Lew \Lambda^2}\int_0^1  Y e^{-\Theta}d\eta \nonumber
\end{align}
where
\begin{equation}
    G_F = \int_0^1 \left[\frac{\mathcal{P}}{2\lambda_*}(1-3\eta^2)Y - \frac{1}{\Lew}\pfr{Y}{X}\right]d\eta, \quad  G_T = \int_0^1 \left[\frac{\mathcal{P}}{2\lambda_*}(1-3\eta^2)\Theta-\pfr{\Theta}{X}\right]d\eta \nonumber
\end{equation}
are just the Taylor dispersion fluxes. As $X\rightarrow -\infty$, $G_F=G_T=1+\gamma \mathcal{P}^2$ as can be verified by substituting the local behaviour of the two-term expansions~\eqref{yfexp}-\eqref{texp} in the above expressions, whereas $X\rightarrow \infty$, $G_F=G_T=0$. Thus, similar to one-dimensional, laminar planar flames, the fluxes entering the reaction zone are made to vanish as they leave that zone, except that the fluxes here are enhanced by Taylor's mechanism from its purely molecular diffusional values. But the more remarkable difference is that the inner transition zone is fundamentally two-dimensional because of the presence of the shear convection term in the governing equations~\eqref{innerY}-\eqref{innerT}. 

When $\mathcal{P}=0$, the problem becomes one-dimensional and $\lambda_*$ disappears from~\eqref{innerY}-\eqref{innerT}. The evanescence of $\lambda_*$ means that the classification based on $\lambda$ (or, equivalently $\lambda_*$) introduced in $\S$1 has no consequence on the burning speed for this case. The burning speed formula for small Peclet numbers reported in the past~\cite{daou2001flame,short2009asymptotic} reveal that the first correction to $\Lambda=1$ is proportional to $\mathcal{P}^2$. Although the Peclet number forcing for the leading-order planar problem is of the order $\mathcal{P}$ (see the left side of~\eqref{innerY} and~\eqref{innerT}) and it can be shown that first corrections to leading order $Y$ and $\Theta$ is also of order $\mathcal{P}$, the burning rate vanishes identically at this order. This interesting asymptotic character for small Peclet numbers was first shown by Zeldovich~\cite{zeldovich1937asymptotic} in a related heat-transfer problem.

\subsection{Asymptotic behaviour of ultra-thick reaction zone regime}
Only in the limit $\lambda_*\rightarrow 0$, full extent of Taylor dispersion effects will be manifested within the reaction zone. Method of solution can be developed in exactly the same manner as in $\S$3.1, but without neglecting the reaction term this time. It is also apparent that the small parameter $\lambda_*$ in~\eqref{innerY}-\eqref{innerT} serves the same purpose as $\ep$ as in~\eqref{gov1}-\eqref{gov3}. It is thus natural to introduce
\begin{align}
    Y &= Y_0(X) + \lambda_* Y_1(X,\eta) + \lambda_*^2 Y_2(X,\eta)+\cdots, \nonumber\\
    \Theta &= \Theta_0(X) + \lambda_* \Theta_1(X,\eta) + \lambda_*^2 \Theta_2(X,\eta)+\cdots, \nonumber\\
   \Lambda &= \Lambda_0 +  \lambda_* \Lambda_1 + \lambda_*^2 \Lambda_2+\cdots \nonumber
\end{align}
into~\eqref{innerY}-\eqref{innerbc3} and solve the problems arising at different orders. The equations satisfied by $Y_0(X)$ and $\Theta_0(X)$ are found to be
\begin{equation}
    \frac{1}{\Lew_{\rm{eff}}}\frac{d^2Y_0}{dX^2}=\frac{d^2\Theta_0}{dX^2}=  \frac{(1+\gamma\mathcal{P}^2)}{2\Lew \Lambda_0^2}Y_0 e^{-\Theta_0}
\end{equation}
which can be reduced to
\begin{equation}
    Y_0 = \Lew_{\rm{eff}}\Theta_0, \quad \frac{d^2\Theta_0}{dX^2} =  \frac{(1+\gamma \mathcal{P}^2)^2}{2\Lambda_0(1+\gamma \mathcal{P}^2\Lew^2)}\Theta_0 e^{-\Theta_0} \label{ultrathick}
\end{equation}
subjected to the conditions $d\Theta_0/dX=-1$ as $X\rightarrow -\infty$ and $d\Theta_0/dX=0$ as $X\rightarrow \infty$. The first integral of this problem provides
\begin{equation}
   \Lambda_0 = \frac{1+\gamma\mathcal{P}^2}{\sqrt{1 + \gamma \mathcal{P}^2\Lew^2}}, \label{asym0}
\end{equation}
the burning-speed formula first derived in~\cite{pearce2014taylor,daou2018taylor}. This formula, which is unaltered when the sign of $\mathcal{P}$ is reversed, carries no information about the direction of the flow. 

The flow-direction dependent formula, obtained by a further pursuit of perturbation analysis for small $\lambda_*$ as summarized in the Appendix, is given by
\begin{align}
    \Lambda = \frac{1+\gamma\mathcal{P}^2}{\sqrt{1 + \gamma \mathcal{P}^2\Lew^2}}\left(1+\lambda_*\frac{\Lambda_1}{\Lambda_0}\right)
\end{align}
where
\begin{align}
     \frac{\Lambda_1}{\Lambda_0} = \frac{7\mathcal{P}}{720}(2-3\Lew \mathcal{I}_1) -  \frac{\Lew\mathcal{I}_2}{1 + \gamma \mathcal{P}^2\Lew^2}\left(\frac{31\mathcal{P}^3\Lew^2}{46200}+ \frac{7\mathcal{P}}{240}\right)  + \frac{2}{3(1 + \gamma \mathcal{P}^2)}\left[ \frac{7\mathcal{P}}{240} (\Lew_{\rm{eff}}-1) \right.   \nonumber \\
       + \left.\frac{31\mathcal{P}^3}{46200}(\Lew^2\Lew_{\rm{eff}}-1)\right] \label{asym1}
\end{align}
and the constants $\mathcal{I}_1\approx 0.4199$ and $\mathcal{I}_2\approx 0.2468$ are defined accordingly as~\eqref{I1} and~\eqref{I2}. Clearly, the appearance of odd powers of $\mathcal{P}$ in $\Lambda_1$ entails information about the flow direction. For the particular case $\Lew=1$, we have
\begin{equation}
    \frac{\Lambda_1}{\Lambda_0}=\frac{7\mathcal{P}}{720}(2-3 \mathcal{I}_1) - \frac{ \mathcal{I}_2}{1+\gamma\mathcal{P}^2} \left(\frac{31\mathcal{P}^3}{46200}+\frac{7\mathcal{P}}{240}\right). \label{asym1Le1}
\end{equation}
The sign of $\Lambda_1$ determines whether $\Lambda$ increases or decreases from its leading value~\eqref{asym0} as $\lambda_*$ increases. It can be shown that $\Lambda_1$ and $\mathcal{P}$ carry same signatures for $\Lew>1.5877$ and when $\Lew<1.1667$, they carry opposite signs. In the intermediate range, $1.1667<\Lew<1.5877$, the sign of $\Lambda_1$ depends on the value of $\mathcal{P}$.

\section{Numerical results}
\subsection{Comparison with the full problem and applicability of the reduced problem}
The reduced problem derived in $\S$3 has used the theories of large activation energy asymptotics and  lubrication. The first theory helped to split the flame problem into an inner reactive and outer non-reactive problems, whereas the second theory allowed to tackle these problems in a simple manner, owing to two reasons. First, in the limit $\ep\rightarrow 0$, thermal expansion effects and therefore the coupling between gas dynamics and the flame arose only as first order corrections. Second, at least for the preheat and post-flame zones, the leading order problems obtained were one-dimensional, as discussed in $\S$3.1. Obviously, these two simplifications are lost when $\ep\sim 1$ or larger.

\begin{figure}[h!]
\centering
\includegraphics[scale=0.6]{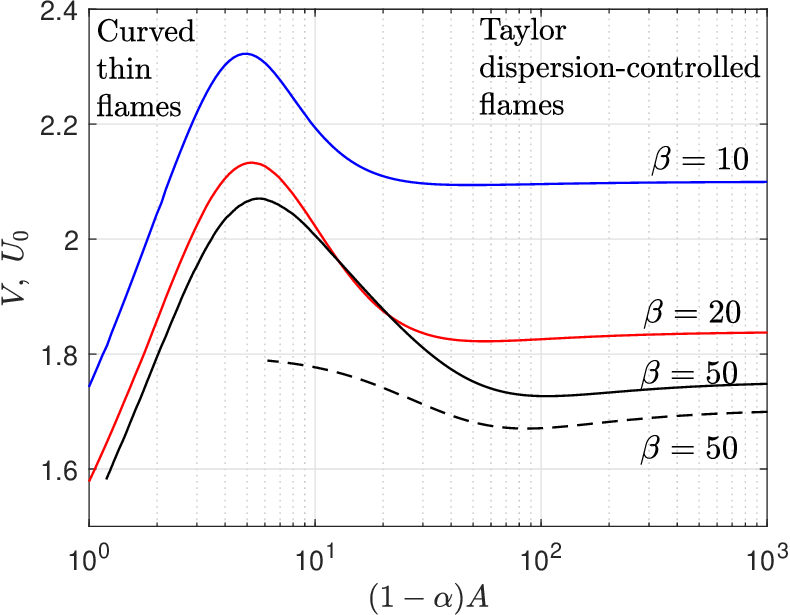}
\caption{Solid lines represent $V$ calculated from~\eqref{full}, whereas the dashed line corresponds to $U_0$, calculated from~\eqref{innerT} as a function of $\lambda_*\sim \beta \epsilon$ and is converted to a function of $\epsilon$ by selecting $\beta=50$. All curves pertain to $\mathcal{P}=10$.}
\label{fig:full}
\end{figure}

To appreciate the range of applicability of the reduced problem that is, in principle, valid for small  values of $1/\beta$ and $\epsilon$, it is essential to understand the behaviour of the full problem for finite values of $\beta$ and $\epsilon$. To that end, we shall assume thermo-diffusive approximation and $\Lew=1$ for the full problem so that solving equation~\eqref{gov5} with the constant-density Poiseuille flow and replacing $y_F=1-\theta$ suffices. There is an arbitrariness in assigning a value for the density in the thermo-diffusive approximation and the following judicious selection for $\varrho$ will allow us to compare results from equation~\eqref{gov5} to the reduced problem of $\S$3.2 in a better manner,
\begin{equation}
    \varrho u \rightarrow \frac{\mathcal{P}}{2\epsilon} (1-3\eta^2), \qquad \varrho V \rightarrow V, \qquad \omega \rightarrow \frac{\beta^2}{2}(1-\theta) e^{-\beta(1-\theta)}
\end{equation}
in which the first and the last substitutions (along with the removal of $\alpha$) use the density at adiabatic flame temperature, because their influences are greatest in the reaction zone, whereas since $V$ is defined with respect to the gas far ahead of the reaction zone, we set $\varrho=1$ for the second term. Equation~\eqref{gov5} thus simplifies to
\begin{equation}
    \left[\frac{\mathcal{P}}{2\epsilon}(1-3\eta^2) +V\right]\pfr{\theta}{\xi} = \pfr{^2\theta}{\xi^2} + \frac{1}{\epsilon^2} \pfr{^2\theta}{\eta^2} + \frac{\beta^2}{2}(1-\theta) e^{-\beta(1-\theta)} \label{full}
\end{equation}
subjected to the conditions $\theta(-\infty,\eta)=\theta(\infty,\eta)-1=0$ and $\pfi{\theta}{\eta}=0$ at $\eta=0,1$ in which we also neglected heat loss effects for simplicity. The problem depends on three parameters, namely, $\mathcal{P}$, $\beta$ and $\epsilon$, whereas for the same case the reduced problem depends on only two because in the latter, only the combination $\beta\ep$ appears.

All numerical computations are performed  using the finite element package COMSOL Multiphysics. Computed values of $V$ from~\eqref{full} for various values of $\beta$ and $\mathcal{P}=10$ are plotted in figure~\ref{fig:full} as functions of \blue{$(1-\alpha) A=\mathcal{P}/\epsilon$}, where $A$ is the flow amplitude discussed in $\S$1. Here, it is worth mentioning how results will modify depending on which parameters are kept constant and which ones are varied. When $\mathcal{P}$ is fixed as we have done in figure~\ref{fig:full}, the non-dimensional flow amplitude varies inversely to $\epsilon$. From the experimental point of view where $L$ is usually fixed, \blue{we can independently alter the flame thickness $\delta_L$ (and consequently modifying $S_L$ since $\delta_L S_L = D_{T,u}$) by adjusting the dilution of the reacting mixture and the flow amplitude $A=U_m/S_L$ by adjusting the flow rate such that $\mathcal{P}= (1-\alpha) U_m L/\delta_LS_L=(1-\alpha) A\epsilon$ remains constant.} In such cases, the burning speed initially increases linearly as $A$ is increased, passes through a maximum when $A\sim 1$ before decreasing it a constant value as $A\rightarrow \infty$; there exists a minimum between the maximum and the final value whose existence depends on the value of $\mathcal{P}$, $\Lew$ and $\beta$. This trend representing the \textit{bending effect}~\cite{ronney1995some} in the laminar case is similar to that reported (as blue lines) in figure 3 of~\cite{daou2018taylor} \blue{and is similar to those curves reported for flame propagation in periodic flows~\cite{kagan2000flame,kagan2005effect}}. In figure~\ref{fig:full}, the burning speed $U_0$ calculated from the reduced problem~\eqref{innerY}-\eqref{innerbc3} with $\Lew=1$ and $\mathcal{P}=10$ is plotted as a dashed curve. It can be seen that results of the full problem and the reduced problem  approach each other when $\beta  $ and $1/\ep \sim A$ is large. More specifically, the numerical results suggest that the reduced model is reasonably accurate when $\beta\gg 1$ and $\ep\ll 1$ with $\lambda_* \leq 10$, approximately.\footnote{We note parenthetically that if $V$ is plotted as a function of $A$ by keeping $\ep$ fixed, then one obtains (not shown herein) curves \blue{without exhibiting the bending effect}, similar to those presented in figure 3 of~\cite{daou2001flame},  figure 8 of~\cite{short2009asymptotic} and (as the green lines of) figure 3 of~\cite{daou2018taylor}; note that in~\cite{short2009asymptotic}, Peclet number is defined with respect to $S_L$ so that keeping their Peclet number fixed is equivalent keeping our $\ep$ constant.}

In summary, the reduced problem discussed in $\S$3.2 for the thick reaction zones describes asymptotically the tail portion of the bending curve. With this understanding, the rest of the numerical results presented in this section will pertain to \blue{the reduced problem described by~\eqref{innerY}-\eqref{innerbc3}, characterizing Taylor dispersion-controlled flames}.

\subsection{The structure of the reaction zone}
\begin{figure}[h!]
\centering
\includegraphics[scale=0.8]{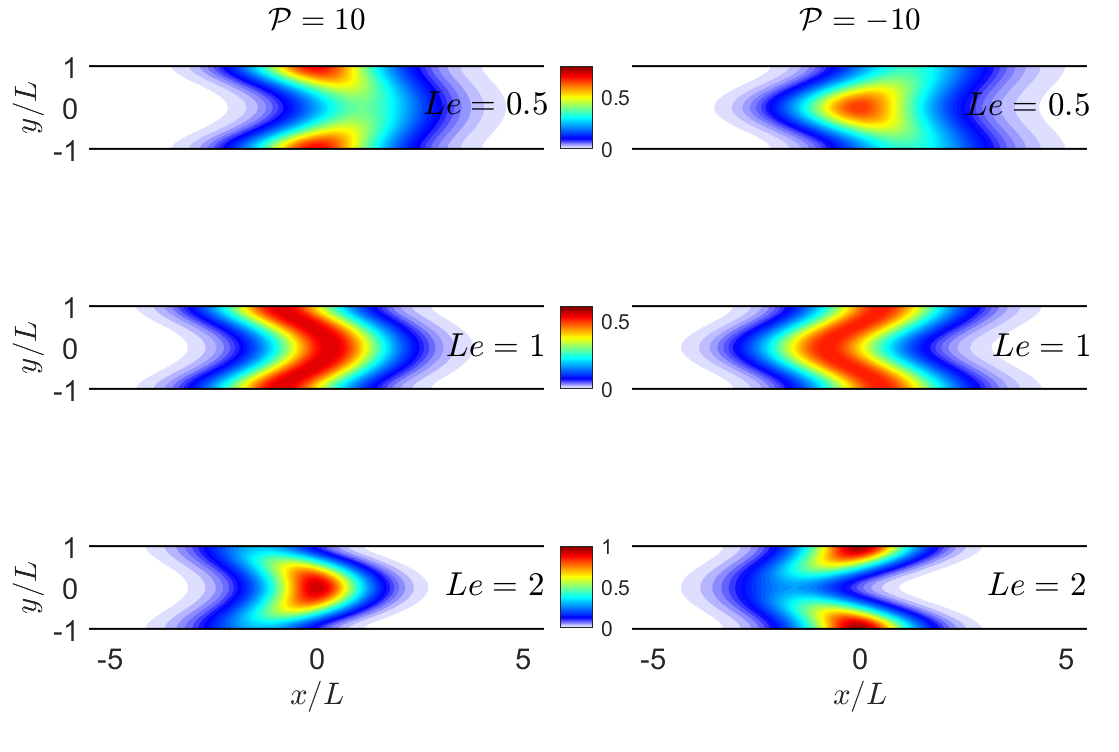}
\caption{Normalized reaction-rate ($\lambda_*^2(1+\gamma \mathcal{P}^2)^2Ye^{-\Theta}/(2\Lew\Lambda^2)$) contours for $\lambda_*=2$. The three left-side plots correspond to $\mathcal{P}=10$ and the right-side figures to $\mathcal{P}=-10$.}
\label{fig:react}
\end{figure}
 The influence of Lewis number on the structure of reaction layer is investigated in Fig.~\ref{fig:react} with $\lambda_*=2$ for $\mathcal{P}=-10,\,10$ and $\Lew=0.5,\,1,\,2$ by exhibiting (normalized) reaction-rate contours. From a broad viewpoint, reaction layers in all cases are curved so that they are concave towards the incoming flow whose direction depends on the sign of $\mathcal{P}$. As is well known for curved flame fronts, preferential diffusion effects can significantly alter the local burning rate; for instance when $\Lew<1$, chemical activity reduces in regions where the front is concave towards the unburnt gas and increases in regions where it is convex to the unburnt gas. Such observations ideally meant for transport processes occurring in the preheat-zone region appear to be valid, as we can infer from Fig.~\ref{fig:react}, for thick reaction layers as well. When $\mathcal{P}$ is further increased, reaction layers are found to exhibit further excursions in the longitudinal direction thereby increasing its (overall) curvature to the extent that we observe chemical activity either near the channel mid-region or near the walls based on the values of $\mathcal{P}$ and $\Lew\neq 1$, similar to that reported in the figure 11 of~\cite{cui2003effects}.

\begin{figure}[h!]
\centering
\includegraphics[scale=0.8]{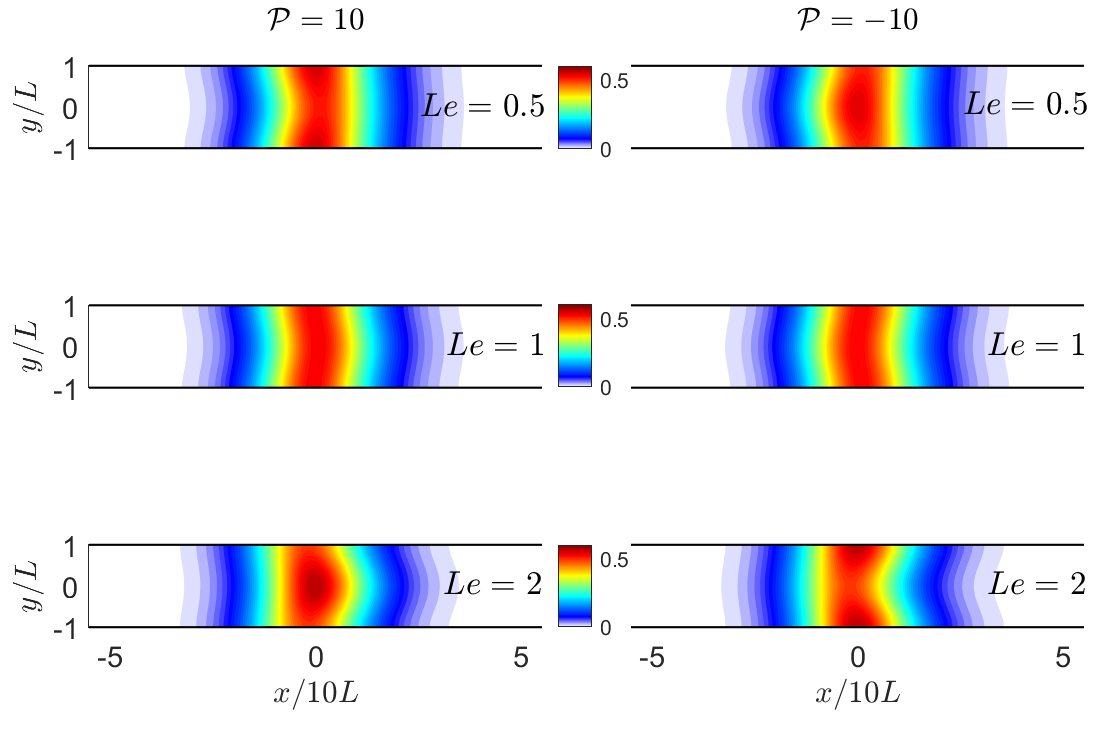}
\caption{Normalized reaction-rate ($\lambda_*^2(1+\gamma \mathcal{P}^2)^2Ye^{-\Theta}/(2\Lew\Lambda^2)$) contours for $\lambda_*=0.2$. The three left-side plots correspond to $\mathcal{P}=10$ and the right-side figures to $\mathcal{P}=-10$.}
\label{fig:react1}
\end{figure}

Fig.~\ref{fig:react1} shows the reaction-rate contours for $\lambda_*=0.2$ selected as a representative case for ultra-thick reaction zone, where in each plot, the $x$-coordinate is stretched by a factor $1/10$ for better visualization. Compared with the previous figure~\ref{fig:react}, here we can see that in all cases, the curving of reaction layers is minimal and therefore for $\Lew\neq 1$, the differential diffusion effects on account of curvature is also negligible. This is a direct consequence of dominance of Taylor's mechanism wherein transverse variations are smoothed out thereby weakening the curvature of concentration and temperature fields. One may logically, thus expect the structure to approach the planar reaction zone but one that is very thick. The primary difference, however, from the planar cases is that the diffusion coefficients are greatly modified by the values of $\mathcal{P}$ and $\Lew$ so that although the structure may resemble a planar reaction layer, its burning rate is altered significantly, as evident from the leading order one-dimensional problem~\eqref{ultrathick}-\eqref{asym0}. 

\subsection{Adiabatic burning speeds}
\begin{figure}[h!]
\centering
\includegraphics[scale=0.65]{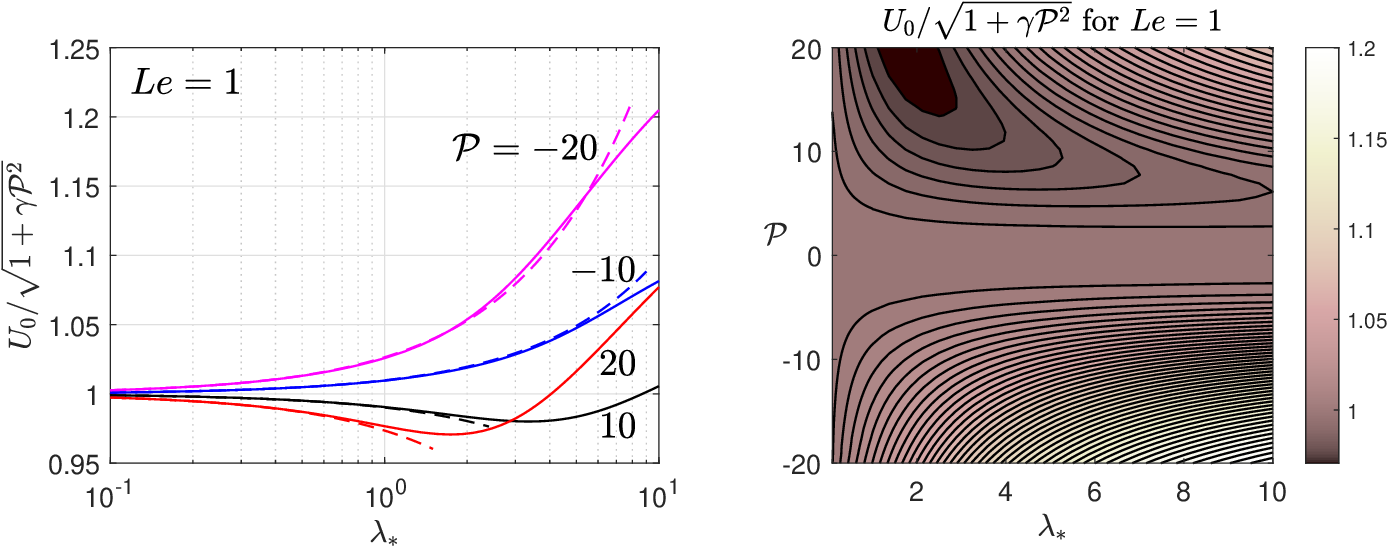}
\caption{Left: Adiabatic burning speeds for $\Lew=1$ as a function of $\lambda_*$ in which the solid lines belong to numerical results and the dashed lines to formula~\eqref{asym1Le1}. Right: Contours of burning speed for a continuous range of $\mathcal{P}$ and $\lambda_*$.}
\label{fig:Uad1}
\end{figure}

\begin{figure}[h!]
\centering
\includegraphics[scale=0.65]{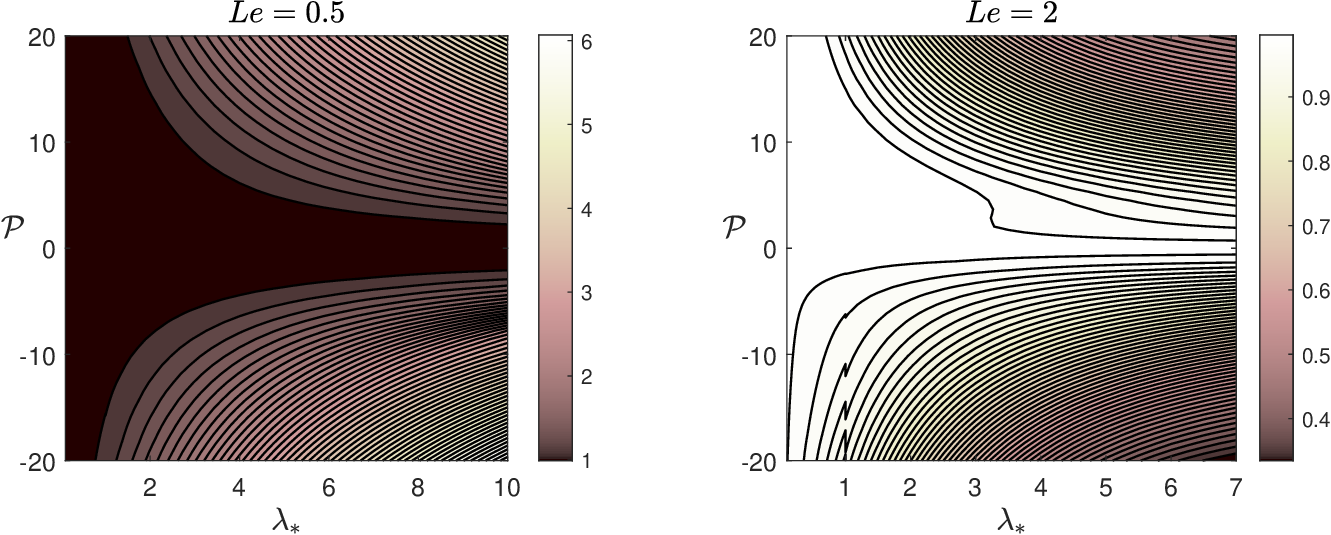}
\caption{Contours of $ U_{0}\sqrt{1+\gamma\mathcal{P}^2\Lew^2}/(1+\gamma\mathcal{P}^2)$ in which the left plot belongs to $\Lew=0.5$ and the right one to $\Lew=2$.}
\label{fig:Uad2}
\end{figure}

\blue{The adiabatic burning speed is simply given by $U_0=\Lambda$, as evident from the last relation in~\eqref{asym1Le1}. According to~\eqref{asym0}, $U_0$ tends to the asymptotic value $(1+\gamma\mathcal{P}^2)/\sqrt{1+\gamma\mathcal{P}^2\Lew^2}$ in the limit $\lambda_*\rightarrow 0$. Scaled by this asymptotic value, $U_0$ is plotted in Fig.~\ref{fig:Uad1} corresponding to $\Lew=1$, and Fig.~\ref{fig:Uad2} corresponding to $\Lew=0.5$ and $\Lew=2$.}

The calculated dependence of burning speed on $\lambda_*$ for $\Lew=1$ is shown in Fig.~\ref{fig:Uad1}. The left plot is graphed for selected values of Peclet number magnitude on a log-scale for $\lambda_*$ whereas the right-side contour plot provides burning speed values for a continuous range of $\mathcal{P}$ and $\lambda_*$ on a normal scale.  In the left-side figure, the asymptotic formula~\eqref{asym1Le1} is as well drawn as dashed lines with which the agreement with the computational results is pretty remarkable for  small $\lambda_*$ where the formula is supposed to be valid. As indicated by that formula for $\Lew=1$, the curves corresponding to opposite signs of $\mathcal{P}$ are symmetric to each other for small $\lambda_*$. The decreasing trend from its value $U_0/\sqrt{1+\gamma\mathcal{P}^2}=1$ for the burning speed with positive values of $\mathcal{P}$ is reversed beyond a particular value of $\lambda_*$ and eventually burning speeds for all $\mathcal{P}$ values increase as $\lambda_*$ increases. Generally, in premixed combustion, thin reaction layers indicate that they consume fuel or, release heat at a faster rate so that the burning speed is large in comparison with thick reaction zones. However, in our case, different levels of thicknesses, characterized by different values of $\lambda_*$, are also subjected to different levels of enhancement of diffusion coefficients by Taylor's mechanism. Any non-monotonic behaviour of the burning rate with reaction-layer thickness can thus be attributed to the interplay between the Taylor's mechanism and curvature effects. The contour plots on the right reveals the full picture of burning-speed dependences for various values of $\mathcal{P}$ and $\lambda_*$.

We now investigate the effects of Lewis numbers on the burning speed for two selected values, $\Lew=0.5,\,2$, in Fig.~\ref{fig:Uad2}. In both cases, for small values of $\lambda_*$, the contours are no longer symmetric about the point $U_{0}=(1+\gamma\mathcal{P}^2)/\sqrt{1+\gamma\mathcal{P}^2\Lew^2}$, a behaviour that is consistent with the formula~\eqref{asym1}. For $\Lew=0.5$, we observe $U_{0}>(1+\gamma\mathcal{P}^2)/\sqrt{1+\gamma\mathcal{P}^2\Lew^2}$ inasmuch as discernible from the figure, whereas in the $\Lew=2$ case, it is other way around, i.e., $U_{0}<(1+\gamma\mathcal{P}^2)/\sqrt{1+\gamma\mathcal{P}^2\Lew^2}$. These trends continue to hold for finite values of $\lambda_*$. In particular, in the second case, the burning speed reduces to values below the adiabatic, planar one-dimensional velocity. The dramatic changes from the planar structure arising due to diffusive-thermal effects controls the burning speed for large values of $\lambda_*$. 

\subsection{\blue{Extinction due to heat loss}}
\blue{The burning speeds $\Lambda$ presented in the previous section represent adiabatic cases whereas for the non-adiabatic cases, the corresponding burning speeds are obtained by solving the last relation in~\eqref{innerscale}, yielding}
\begin{equation}
    \blue{ U_0 = \sqrt{\frac{-2\kappa(1+\gamma\mathcal{P}^2)}{W_0[-2\kappa(1+\gamma\mathcal{P}^2)/ \Lambda^2]}}}
\end{equation}
\blue{where $W_0$ is the principal branch of the Lambert W function. The Lambert W function $W(x)$ is real only when its argument $x\geq -1/e$ and consequently $U_0$ exists only if $\kappa \leq \kappa_{\mathrm{ext}}$, where $\kappa_{\mathrm{ext}}$ is the value of $\kappa$ corresponding to the extinction point. This extinction value and the corresponding burning speed are given by }
\begin{equation}
    \blue{\kappa_{\mathrm{ext}} = \frac{ \Lambda^2}{2e(1+\gamma\mathcal{P}^2)}\approx  \frac{ 0.1839\, \Lambda^2}{1+\gamma\mathcal{P}^2} , \qquad U_{0,\mathrm{ext}} = \frac{ \Lambda}{\sqrt{e}}\approx  0.6065\,\Lambda.} \label{kext}
\end{equation}
\blue{in which $\Lambda$ is a function of $\lambda_*$, $\mathcal{P}$ and $\Lew$.}

\blue{In the ultra-thick reaction zone regime ($\lambda_*\rightarrow 0$), we can replace $\Lambda$ with its leading value~\eqref{asym0} to find}
\begin{equation}
     \blue{\kappa_{\mathrm{ext}}=  \frac{1+\gamma\mathcal{P}^2}{2e(1+\gamma\mathcal{P}^2\Lew^2)} \quad \text{and} \quad U_{0,\mathrm{ext}}= \frac{1+\gamma\mathcal{P}^2}{\sqrt{e(1+\gamma\mathcal{P}^2\Lew^2)}}.}
\end{equation}
\blue{These results for near equidiffusional flames with $\Lew\approx 1$ are obtained in~\cite{daou2022flame}. In the absence of the shear flow, i.e., when $\mathcal{P}=0$, the above formulas reduce to $\kappa_{\mathrm{ext}}=1/2e$ and $U_{0,\mathrm{ext}}=1/\sqrt{e}$. However, in the presence of a strong shear flow, $\mathcal{P}\gg 1$, we obtain}
\begin{equation}
     \blue{\kappa_{\mathrm{ext}}=  \frac{1}{2e\Lew^2} \quad \text{and} \quad U_{0,\mathrm{ext}}= \frac{\sqrt{\gamma}\mathcal{P}}{\sqrt{e}\Lew}.}
\end{equation}
\blue{The first formula implies that lighter fuels can sustain combustion with relatively larger heat losses. For instance, the quenching limit of hydrogen flames ($\Lew\approx 0.3$) in the presence of a strong shear flow should be about $11$ times greater than its value in the absence of flow.}

\blue{As we can see from figure~\eqref{fig:Uad2}, for a fixed value of $\mathcal{P}$, the adiabatic burning speed $\Lambda$ increases as $\lambda_*$ increases for $\Lew=0.5$ and decreases as $\lambda_*$ increases for $\Lew=2$. This implies that $\kappa_{\mathrm{ext}}$ also increases or decreases according to the first formula in~\eqref{kext}. Therefore, we may state that, generally speaking, Taylor dispersion extends the flammability limits for lighter fuels and constricts them for heavier fuels. Such effects of unidirectional shear flows on the flammability limits do not appear to have been recognized in previous publications.}

\section{Implications of the model problem for turbulent combustion in the distributed reaction zone regime}
According to Damk{\"ohler}'s second hypothesis~\cite{damkohler1940einfluss}, the effect of small scale turbulence  is to enhance the effective diffusion coefficients and enhance the effective burning speed $S_T$, without changing the flame structure. Therefore, using formula~\eqref{SL}, we may write
\begin{equation}
    \frac{S_T}{S_L} = \sqrt{\frac{\Lew_{\rm{tur}} D_{\rm{T,tur}}}{\Lew\, D_T}}  \label{Damo}
\end{equation}
where $D_{T,\rm{tur}}$ is the turbulent thermal diffusivity and $\Lew_{\rm{tur}}=D_{T,\rm{tur}}/D_{F,\rm{tur}}$ is the turbulent Lewis number, involving the turbulent fuel diffusion coefficient $D_{F,\rm{tur}}$. Although it is often commonly assumed $\Lew_{\rm{tur}}= 1$, many recent studies~\cite{aspden2011characterization,rieth2022enhanced,cai2022turbulent} point to a dependence on the Lewis number for highly turbulent flames. In particular, the recent experimental study~\cite{cai2022turbulent} leads to the correlation formula
\begin{equation}
    \frac{S_T}{S_L} \sim \frac{\sqrt{\Rey}}{\Lew} \sim \frac{\Rey_\lambda}{\Lew} \label{new}
\end{equation}
in the distributed reaction zone regime. Here $\Rey=u'l/\nu$ is the turbulent Reynolds number involving the rms fluctuating velocity $u'$ and the integral scale $l$. Note that in~\eqref{new}, we have  introduced the Reynolds number $\Rey_\lambda=u'\lambda/\nu$ based on the Taylor microscale $\lambda\sim l/\sqrt{\Rey}$~\cite{pope2000turbulent}. Formulas~\eqref{Damo} and~\eqref{new} imply that
\begin{equation}
    \Lew_{\rm{tur}}= \frac{1}{\Lew} \qquad \text{and} \qquad \frac{D_{T,\rm{tur}}}{D_T}\sim  \Rey. \label{Letur}
\end{equation}

It is interesting to compare now the results above with the results of our laminar model problem.
 To this end, we turn to formula~\eqref{Leeff} for the effective Lewis number $\Lew_{\rm{eff}}$ and formula~\eqref{asym0} for scaled effective burning speed $S_T/S_L= U_{ad} + O(\ep^2)$, which we rewrite for convenience 
 \begin{equation}
   \Lew_{\rm{eff}} = \Lew \frac{1+\gamma\mathcal{P}^2}{1+\gamma\mathcal{P}^2\Lew^2} \qquad \text{and} \qquad \frac{S_T}{S_L} \sim \frac{1+\gamma\mathcal{P}^2}{\sqrt{1 + \gamma \mathcal{P}^2\Lew^2}}. \label{ours0}
\end{equation}
For $\mathcal{P}\gg 1$, we thus obtain
\begin{equation}
   \Lew_{\rm{eff}}\sim \frac{1}{\Lew} \qquad \text{and} \qquad \frac{S_T}{S_L} \sim \frac{\mathcal{P}}{\Lew}. \label{ours}
\end{equation}
It is interesting to note that the dependence of $\Lew_{\rm{eff}}$ given by~\eqref{ours} (pertaining to our laminar model) and the dependence of $\Lew_{\rm{tur}} $ given by~\eqref{Letur} (pertaining to the turbulent case) on the Lewis number $\Lew$ are identical. Similarly, the dependence of $S_T/S_L$ on the Lewis number, being the same in the laminar case~\eqref{ours} and turbulent case~\eqref{new} is remarkable. This alludes that a diffusion enhancement mechanism generalizing Taylor dispersion mechanism may be operative for highly turbulent flame. This aspect deserves further investigation in line with the studies~\cite{majda1999simplified,shende2022nonlocal} addressing diffusion enhancement models in flow fields.

Finally we note that dependence of $S_T/S_L$ on the Reynolds number is tricky to reconcile in the laminar and turbulent cases if $\mathcal{P}$ is identified with a  Reynolds number. This is perhaps not surprising since our laminar model comprises a single length scale, whereas the turbulent problem comprises several length scales. Interestingly however, by comparing~\eqref{new} and~\eqref{ours}, it is seen that $\mathcal{P}$ may be viewed as the laminar analogue of $\Rey_\lambda$, rather than $\Rey$.

\section{Concluding remarks}

A primary objective of this study is to develop a model illustrating the structure emerging when a shear flow interacts with a thick reaction zone. One of the key parameter that controls various physical phenomena is the ratio of channel half-width to reaction zone thickness, $\lambda_*$, defined as in~\eqref{lambdaT}. When this parameter is small, Taylor dispersion is found to be predominant with the flame front experiencing negligible distortions and curvature effects. This leads, for small values of $\lambda_*$, to enhanced diffusion resulting in an enhanced burning speed, determined by 
the analytical expression~\eqref{asym1}. In contrast, when $\lambda_*$ is large, Taylor-dispersion effects are found to be negligible while flame distortion significantly influences the effective burning speed. In the latter case, preferential diffusion effects are also significant when $\Lew\neq 1$ resulting in familiar features such as the tip opening of a curved front. The model derived given by~\eqref{innerY}-\eqref{innerbc3}, involving three nondimensional parameters $\mathcal{P},\,\Lew$ and $\lambda_*$, is obtained from the reactive Navier-Stokes equations~\eqref{gov1}-\eqref{BC4} using large activation energy asymptotics and lubrication theories, involving the seven parameters $\ep,\,\Pra,\,\Lew,\,\Pec,\,\beta,\,\alpha$ and $\kappa$. 

\blue{The extinction heat loss parameter in Taylor dispersion-controlled flames is found to be multiplied by a factor $1/\Lew^2$ in comparison with those cases where there is no shear flow. The implication is therefore flow-induced diffusion extends the flammability limit for lighter fuels and constricts for heavier fuels. It may therefore be plausible to expect that extended flammability limits may be obtained for thick reaction zone hydrogen flames. It is of interest therefore to consider a detailed~\cite{carpio2020near} or reduced~\cite{fernandez2019one} chemistry description to improve quantitative predictions of such flames in near-extinction critical conditions.}

In spirit, the problem~\eqref{innerY}-\eqref{innerbc3} may be loosely viewed as the laminar analogue of a  premixed flame problem in the distributed reaction zone regime in turbulent combustion~\cite{peters2000turbulent}. This model predicts an effective Lewis number and an effective burning speed formulas, given by~\eqref{ours0}. Theses formulas imply for $\mathcal{P}\gg 1$ a dependence on $\Lew$, given by~\eqref{ours}, which coincides with the results of a recent experimental study~\cite{cai2022turbulent} carried out for highly turbulent jet flames. The peculiar $1/\Lew$ dependence in our model emerges due to Taylor dispersion and therefore we speculate that this dependence may be attributed to a similar but more general mechanism in the highly turbulent case, rather than to a coupling between differential diffusion and the flame curvature. Thus, our model suggests that the role of differential diffusion is not the same in large scale and small scale turbulence, although its effect is felt in both cases.

The qualitative prediction from a simple parallel flow field is remarkable enough to motivate future studies to consider the interactional effects between a thick reaction zone and prescribed flow fields. Such flows may include spatio-temporal periodic flows, multiple scale flow fields and others used in previous investigations such as~\cite{daou2007flame,kagan2000flame,bourlioux2000elementary,kagan2005effect,majda1999simplified}.

\section*{Acknowledgements}
This research was supported by the UK EPSRC through grant EP/V004840/1.

\bibliographystyle{tfq}
\bibliography{interacttfqsample}

\section*{Appendix: Asymptotic analysis for ultra-thick reaction zone regime}
The main aim of this appendix is derive equation~\eqref{asym1}.

\subsection*{Derivation of Taylor's dispersion equations in the presence of chemical reaction}
The problems at first two orders are identical to the preheat zone problems. The solutions at first order are
\begin{align}
    Y_1 &= Y_{1a}(X) + \mathcal{P}\Lew \frac{dY_0}{dX}\left(\frac{\eta^2}{4}-\frac{\eta^4}{8}\right), \nonumber \\
    \Theta_1 &= \Theta_{1a}(X) + \mathcal{P} \frac{d\Theta_0}{dX}\left(\frac{\eta^2}{4}-\frac{\eta^4}{8}\right). \nonumber
\end{align}
The reaction term enters in the second order problem as follows
\begin{align}
    \frac{\mathcal{P}}{2}(1-3\eta^2) \pfr{Y_1}{X} = \frac{1}{\Lew}\left(\frac{d^2Y_0}{dX^2}+\pfr{^2Y_2}{\eta^2}\right) - \frac{(1+\gamma\mathcal{P}^2)^2}{2\Lew \Lambda_0^2}Y_0 e^{-\Theta_0}, \nonumber\\
     \frac{\mathcal{P}}{2}(1-3\eta^2) \pfr{\Theta_1}{X} = \frac{d^2\Theta_0}{dX^2}+\pfr{^2\Theta_2}{\eta^2} - \frac{(1+\gamma\mathcal{P}^2)^2}{2\Lew \Lambda_0^2}Y_0 e^{-\Theta_0} \nonumber
\end{align}
Integrating these equations using the lateral boundary conditions, we obtain
\begin{align}
    Y_2 &= Y_{2a}(X) + \mathcal{P}\Lew \frac{dY_{1a}}{dX}\left(\frac{\eta^2}{4}-\frac{\eta^4}{8}\right) + \mathcal{P}^2\Lew^2 \frac{d^2Y_0}{dX^2}\left(\frac{3\eta^8}{896} - \frac{7\eta^6}{480} + \frac{\eta^4}{96} + \frac{\gamma\eta^2}{2}\right), \nonumber\\
    \Theta_2 &= \Theta_{2a}(X) + \mathcal{P} \frac{d\Theta_{1a}}{dX}\left(\frac{\eta^2}{4}-\frac{\eta^4}{8}\right) + \mathcal{P}^2 \frac{d^2\Theta_0}{dX^2}\left(\frac{3\eta^8}{896} - \frac{7\eta^6}{480} + \frac{\eta^4}{96} + \frac{\gamma\eta^2}{2}\right). \nonumber
\end{align}

The basic equations emerging at third order are
\begin{align}
    \frac{\mathcal{P}}{2}(1-3\eta^2) \pfr{Y_2}{X} = \frac{1}{\Lew}\left(\frac{d^2Y_1}{dX^2}+\pfr{^2Y_3}{\eta^2}\right) + \frac{(1+\gamma\mathcal{P}^2)^2}{2\Lew \Lambda_0^2}\left(Y_0\Theta_1-Y_1 + \frac{2\Lambda_1}{\Lambda_0}Y_0 \right) e^{-\Theta_0}, \nonumber\\
     \frac{\mathcal{P}}{2}(1-3\eta^2) \pfr{\Theta_2}{X} = \frac{d^2\Theta_1}{dX^2}+\pfr{^2\Theta_3}{\eta^2} + \frac{(1+\gamma\mathcal{P}^2)^2}{2\Lew \Lambda_0^2}\left(Y_0\Theta_1-Y_1 + \frac{2\Lambda_1}{\Lambda_0}Y_0 \right) e^{-\Theta_0}. \nonumber
\end{align}
Integrating the above equations across the channel, we can derive a system of linear equations
for the unknown $\vect x=[Y_{1a}\,\,\,\, \Theta_{1a}]^T$ that is expressed in a matrix form $\vect A\vect x=\vect f$ where
\begin{equation}
    \vect A = \begin{bmatrix}
    \frac{1+\gamma\mathcal{P}^2Le^2}{Le}\frac{d^2}{dX^2} -  \frac{ (1+\gamma\mathcal{P}^2)^2e^{-\Theta_0}}{2\Lew \Lambda_0^2}
    &  \frac{(1+\gamma\mathcal{P}^2)^2Y_0e^{-\Theta_0}}{2\Lew \Lambda_0^2}\\
    - \frac{ e^{-\Theta_0}}{2\Lew \Lambda_0^2}
    & (1+\gamma\mathcal{P}^2)\frac{d^2}{dX^2} +  \frac{(1+\gamma\mathcal{P}^2)^2 Y_0e^{-\Theta_0}}{2\Lew \Lambda_0^2}
    \end{bmatrix} \nonumber
\end{equation}
and
\begin{equation}
    \vect f = \begin{bmatrix}
     -\frac{d^3Y_0}{dX^3} \left(\frac{31\mathcal{P}^3Le^2}{23100}+\frac{7\mathcal{P}}{120}\right)   + \frac{7\mathcal{P}(1+\gamma\mathcal{P}^2)^2e^{-\Theta_0}}{240Le \Lambda_0^2}\left(Le\frac{dY_0}{dX} - Y_0\frac{d\Theta_0}{dX}\right) -(1+\gamma\mathcal{P}^2)^2\frac{\Lambda_1Y_0 e^{-\Theta_0}}{Le \Lambda_0^3}
   \\
 -\frac{d^3\Theta_0}{dX^3} \left(\frac{31\mathcal{P}^3}{23100}+\frac{7\mathcal{P}}{120}\right)   + \frac{7\mathcal{P}(1+\gamma\mathcal{P}^2)^2e^{-\Theta_0}}{240Le \Lambda_0^2}\left(Le\frac{dY_0}{dX} - Y_0\frac{d\Theta_0}{dX}\right) -(1+\gamma\mathcal{P}^2)^2\frac{\Lambda_1Y_0 e^{-\Theta_0}}{Le \Lambda_0^3}
    \end{bmatrix} \nonumber
\end{equation}
A suitable inner product involving two vectors, say, $\vect x=[P\,\,\,\,\Phi]^T$ and $\vect y =[Q\,\,\,\,\Psi]^T$ may be defined by
\begin{equation}
    \langle \vect x,\vect y\rangle = \int_{-\infty}^\infty  (PQ + \Phi \Psi)\, dX.  \nonumber
\end{equation}
The solvability condition for the linearized problem can be derived from the Fredholm alternative which states that $\vect A \vect x = \vect f$ has a solution only if $\vect \langle \vect f, \vect x_h\rangle=0$ where $\vect x_h$ the homogeneous solution satisfies $\vect A^T\vect x_h=0$.

\subsection*{Non-trivial homogeneous solution}
Before solving for the homogeneous solution, let us note that considerable simplifications of terms appearing in the linear system occur by using the relations $ \Lambda_0=(1+\gamma \mathcal{P}^2)/\sqrt{1+\gamma\mathcal{P}^2\Lew^2}$, $Y_0=\Lew_{\rm{eff}}\Theta_0$ and then expressing derivatives of $\Theta_0$ in terms of $\Theta_0$ using  $d\Theta_0/dX=-\sqrt{1-(\Theta_0+1)\exp(-\Theta_0)}$, $d^2\Theta_0/dX^2=\frac{1}{2}\Theta_0 \exp(-\Theta_0)$ and $d^3\Theta_0/dX^3=\frac{1}{2}\sqrt{1-(\Theta_0+1)\exp(-\Theta_0)}(\Theta_0-1)\exp(-\Theta_0)$. 

The equations $\vect A^T\vect x_h=0$ defining the homogeneous solution $\vect x_h=[Y_{1a,h}\,\,\,\,\Theta_{1a,h}]^T$ are given by
\begin{align}
    \frac{d^2Y_{1a,h}}{dX^2} - \frac{1}{2}(Y_{1a,h}+\Theta_{1a,h}) e^{-\Theta_0} =0, \nonumber\\
    \frac{d^2\Theta_{1a,h}}{dX^2} + \frac{1}{2}(Y_{1a,h}+\Theta_{1a,h}) \Theta_0 e^{-\Theta_0} =0. \nonumber
\end{align}
Adding the above two equations and integrating, we obtain $Y_{1a,h}+\Theta_{1a,h}=d\Theta_0/dX$. Substituting this result into the second equation, we get
\begin{equation}
   \frac{d^2\Theta_{1a,h}}{dX^2} + \frac{1}{2}\frac{d\Theta_0}{dX} \Theta_0 e^{-\Theta_0} =0. \nonumber
\end{equation}
Integrating this last equation using the condition $\Theta_{1a,h}=0$ at $\Theta_0=0$, the solution can be expressed as
\begin{equation}
    \Theta_{1a,h}(\Theta_0) = \frac{1}{2}\int_0^{\Theta_0}\sqrt{1-(t+1)e^{-t}}\, dt. \nonumber
\end{equation}
Thus, the non-trivial homogeneous solution is given by
\begin{equation}
    \vect x_h =  \begin{bmatrix}
           d\Theta_0/dX-\Theta_{1a,h} \\
          \Theta_{1a,h}
         \end{bmatrix} \nonumber
\end{equation}

\subsection*{The solvability condition}
The unknown constant $\Lambda_1$ is now determined by imposing the solvability condition $\vect \langle \vect f, \vect x_h\rangle=0$. Since all the terms in both $\vect f$ and $\vect x_h$ are expressible in terms of $\Theta_0$, the inner product $\vect \langle \vect f, \vect x_h\rangle$ is easily evaluated by changing the integration variable from $-\infty <X<\infty$ to $0\leq \Theta_0<\infty$. Doing this, we get the following formula
\begin{align}
    \frac{\Lambda_1}{\Lambda_0} = \frac{7\mathcal{P}}{720}(2-3\Lew \mathcal{I}_1) -  \frac{\Lew\mathcal{I}_2}{1 + \gamma \mathcal{P}^2\Lew^2}\left(\frac{31\mathcal{P}^3\Lew^2}{46200}+ \frac{7\mathcal{P}}{240}\right) \nonumber \\
      + \frac{2}{3(1 + \gamma \mathcal{P}^2)}\left[\frac{31\mathcal{P}^3}{46200}(\Lew^2\Lew_{\rm{eff}}-1)+ \frac{7\mathcal{P}}{240} (\Lew_{\rm{eff}}-1)\right] \nonumber
\end{align}
where 
\begin{align}
    \mathcal{I}_1 &= \int_0^\infty e^{-t}\sqrt{1-(t+1)e^{-t}}\,dt,  \label{I1}\\ 
    \mathcal{I}_2 &= \int_0^\infty (t-1)e^{-t}\sqrt{1-(t+1)e^{-t}}\,dt. \label{I2}
\end{align}
From here, equation~\eqref{asym1} follows.

\end{document}